\documentclass[a4paper,12 pt]{article}
\usepackage[utf8]{inputenc}
\usepackage{amsfonts, etoolbox, gensymb, graphicx, nicefrac, natbib, hyperref}
\usepackage[paperwidth=8.5in, paperheight=11in, margin=0.5in,headheight=64pt,marginparwidth=0pt,textheight=650pt]{geometry}

\title{Mr. Bayes and the classics: a suggested interpretation}
\author{
   Marcio A. Diniz\thanks{Department of Statistics, Federal University of São Carlos, Rod. Washington Luis, km 235, São Carlos, Brazil. Email: marciodiniz@ufscar.br. Corresponding author.}
   \and
   David R. Bellhouse\thanks{Department of Statistical and Actuarial Sciences, University of Western Ontario, 1151 Richmond Street, London, Canada. Email: bellhouse@stat.uwo.ca.}}
\date{}

\begin{document}

\maketitle

\begin{abstract}

The main hypothesis about Thomas Bayes’s intentions to write his famous {\it Essay} on probability is that he wanted to refute the arguments of David Hume against the reliability of the occurrence of miracles, published in 1748. 
In this paper we argue that it was not Bayes’s intention to rebut Hume but that his interest on the ``inverse problem'' came about as result of his study of the second edition of Abraham De Moivre’s book, {\it The Doctrine of Chances}, published in 1738.

A possible communication of Bayes’s breakthrough might have annoyed De Moivre, leading to a response written for Bayes in the third edition of De Moivre’s book, published in 1756.  
Among other points, the response claims that De Moivre was the first to solve the mentioned inverse problem. Under this perspective Richard Price’s letter, written as preface to Bayes’s essay, has a new interpretation, appearing also as a defense of Bayes premiership on a satisfactory or proper solution.

\

\noindent
{\bf keywords}: law of large numbers, inverse probability, De Moivre, Thomas Bayes, Bayesian inference
\end{abstract}

\section{Introduction}

The motivations that led Thomas Bayes (c. $1701-1761$) to work on his now famous essay on probability (\cite{bayes1763a}) have been searched by historians for at least the last four decades.\footnote{See \cite{stigler1983, dale1988, edwards1986}.}
One hypothesis is that Bayes, a Nonconformist minister, wanted to refute the arguments of the sceptic philosopher David Hume ($1711-1776$) against the reliability of the occurrence of miracles, published in 1748 (Hume (1748)). 
This hypothesis finds support in the prefatory letter to Bayes’s essay, written by his friend Richard Price ($1723-1791$), also a Nonconformist minister. In fact, Price used the results proved by Bayes as an argument to reply to Hume (\cite{price1767}), later engaging in correspondence with the philosopher.

Our goal is to suggest a different perspective on the same story. 
We will argue that it was not Bayes’s intention to rebut Hume but that his interest on the ``inverse problem'' came about as result of his study of the second edition of Abraham De Moivre’s book, {\it The Doctrine of Chances} (1738). 
A possible communication, without details, of Bayes’s breakthrough might have annoyed De Moivre since this would not be the first time his work was expanded or corrected by Bayes. 
This annoyance may have led to a response written for Bayes in the third edition of De Moivre’s book, published in 1756. 
We advocate that this response was not written by De Moivre himself but by the editor of the third edition of his book, Patrick Murdoch (d. 1774).
Murdoch's response may have been motivated, or asked for, by De Moivre.
Among other points, the response claims that De Moivre was the first to solve the mentioned inverse problem. Under this perspective Price’s prefatory letter has a new interpretation, appearing also as a defense of Bayes premiership on a satisfactory or proper solution. 

The following section reviews the literature to explain what was the ``inverse problem'' and how the scholars approached it before Bayes. 
This will lead us to a possible solution mentioned by David Hartley in a book published in 1749, without naming who suggested it. 
We review the evidence pointing to Bayes as the one responsible for this anonymous reference. 
Section 3 brings the arguments indicating that it was not Bayes’s intention to refute Hume and in Section 4 we describe how Bayes became interested in probability, eventually approaching the inverse problem sometime between 1747 and 1750. 
Section 5 presents the evidence suggesting that Murdoch was the author of the text added in De Moivre (1756) to, possibly, address Bayes. 
We conclude by reviewing Price’s prefatory letter to Bayes’s essay. 

\section{Context: Bernoulli, De Moivre and Hartley}

During the seventeenth and first half of the eighteenth centuries, the typical approach to a probability problem was to assume that the probabilities of some simple events are known. From these basic assumptions, probabilities of more complex events were calculated. By 1720, such calculations were done by Jakob Bernoulli ($1655-1705$), Pierre R\'emond de Montmort ($1678-1719$) and De Moivre ($1667-1754$), among others. Later scholars named these methods ``direct.'' The great leap forward in the mid eighteenth century would be what David Hartley ($1705-1757$) called the ``inverse Problem.'' In the context of eighteenth-century probability calculations this can be stated as: Given the observed relative frequency of some event in a finite number of similar trials, what can be said about the underlying probability of this event?

%The inverse problem was applied in a practical way without theoretical foundation. For example, based on data collected from the City of Breslaw, \cite{halley1693} estimated the number of people alive at each age in that city. From this population table he estimated various survival probabilities, treating these estimates as the true values. From these estimated survival probabilities, he estimated the value of life annuities at various ages. Again, he treated these estimates as the true values.

Sometime before 1690, Bernoulli proved the first limit theorem of probability theory.
His result---a particular case of what we now call weak Law of Large Numbers---was posthumously published in {\it Ars Conjectandi} (1713). Bernoulli's case of the law of large numbers was also a result in direct probabilities. 
It shows how to compute the probability that the relative frequency of repeated trials, all with a given probability of ``success'', will deviate from this underlying probability when the number of trials increases.\footnote{See the Appendix for the formal statement of Bernoulli's result and its inverse use.} Bernoulli realized that the result could be applied  to situations where events were no longer considered equiprobable, opening a wide range of applications to probability theory outside the scope of games of chance. 

\cite{daston1988} argues that passages from Part IV of \cite{bernoulli1713} and, more explicitly, Jakob's correspondence with Gottfried Leibniz ($1646-1716$), indicate that Bernoulli viewed the inverse use of his theorem as a way to solve\footnote{A formal solution of the inverse problem, as stated above, requires the computation of a conditional probability. See the Appendix for the differences between Bernoulli's law, its inverse and Bayes's theorem.} the inverse problem.
Leibniz thought the solution was not satisfactory and raised some points against it.
Bernoulli's rebuttals to Leibniz were eventually published in the {\it Ars  Conjectandi}.\footnote{See \citealp[p.~227]{bernoulli1713} and \citealp[p.~329-30]{bernoulli2006}.} 
\cite{hacking1971,hacking1975} and \cite{dale1988} say that even though Bernoulli’s intentions can be inferred from his book and letters, he did not claim to have proved the explicit result. 
In any case, as \cite{dale1988} comments, at Jakob Bernoulli’s death ``one was left with a careful proof of the direct theorem and a hint at the inverse result.''

De Moivre arrived at his normal approximation to the binomial due to the answer he gave to a challenge problem proposed to him by Alexander Cuming in 1721. 
Comments from \cite{moivre1738} on this approximation show that he also touched on the inverse uses of Benoulli's limit theorem.

De Moivre's work on the subject  was firstly discussed in Book V of the {\it Miscellanea Analytica}, \cite{moivre1730}, where he transcribed several paragraphs from \cite{bernoulli1713}.
To his conclusions on the normal approximation to binomial probabilities, he added further work refining Bernoulli's computations in a pamphlet called {\it Approximatio ad Summam Terminorum Binomii $\overline{a+b}|^n$ in Seriem expansi}, published at his own expense in 1733 and distributed to a few friends.

The second edition of {\it The Doctrine of Chances}, published in 1738, brings the English translation of the {\it Approximatio} exactly after the problems proposed by Cuming in 1721.\footnote{See \citealp[p.~231,233]{moivre1738}.}
After the translation, De Moivre added a text with numerical illustrations of the main results, finishing with a brief defense of the design argument.\footnote{See \citealp[p.~243]{moivre1738}.} 

\cite{dale1988} and \cite{edwards1986} argue that in the {\it Approximatio} De Moivre revealed his intention to use Bernoulli's theorem to solve the inverse problem.
\cite{edwards1986} mentions other elements to confirm this hypothesis, the most compelling one being the Advertisement of the second edition, published in the first pages of the book.
In it De Moivre gives a list of reasons showing why the second edition was better than the first one.
The 13th reason is that the second edition brings

\

\begin{center}
\begin{minipage}[c]{0.8\textwidth}
\begin{small}
The Solution of a Question which cannot fail of interesting the Reader; it is, What reasonable Conjectures may be derived from Experiments, or what are the Odds that after a certain number of Experiments have been made concerning the happening or failing of Events, the Accidents of Contingency will not afterwards vary from those of Observation beyond certain Limits; which leads naturally to this Consequence, that altho' Chance has very great Influence on some Events, yet that it very little disturbs those which in their natural Institution were designed to happen according to fixt Laws, Chance vanishing as it were at long run in respect to the Constancy and Regularity of Order.
\end{small}
\end{minipage}
\end{center}

\

The text asks if we can reasonably conjecture based on past experience.
The writer of the advertisement answers that we can, saying that while single events are unpredictable, there is stability and order in the long run.
Thus, the advertisement says the 1738 edition provided an answer to the inverse problem.\footnote{In 1736 the {\it Bibliothèque raisonnée des ouvrages des savans de l'Europe}, a French journal that had a section with reports on new publications coming out of London, advertised the second edition of {\it The Doctrine of Chances} with almost exactly the same text.
Compare the paragraph given above with the following one, from the {\it Bibliothèque raisonnée}, volume 16, p. $218-219$: ``Enfin, on y trouvera la solution de cette Question curieuse \& interessante: Qu'elles conjectures peut-on raisonnablement sonder sur l'Expérience? ou, Qu'elle probabilité y a-t-il qu'après avoir trouvé par un certain nombre d'expériences, que des événemens sont ou ne sont pas arrivez les accidens de la Contingence ne s'éloigneront pas après cela de ceux qu'on a observez au delà de certaines limites. Ce qui conduit naturellement à cette conséquence; Que quoi que le Hazard influe beaucoup sur certains evénemens cependant il dérange fort peu ceux qui ont été d'abord déterminez à arriver selon des Loix fixes \& constantes; le Hazard s'évanouissiant pour ainsi dire, à la longue, par rapport à la stabilité \& au retour de l'ordre.'' .
\cite{bellhouse2011} conjectured that the author of the 1736 text was Pierre Des Maizeaux. 
He was one of the active writers for the {\it Bibliothèque raisonnée}.
Notwithstanding, the similitude of both texts indicate that they were probably written by De Moivre himself. In the 1738 text De Moivre used the first person. See \citealp[p.~xii$-$xiv]{moivre1738}.} 

\subsection{David Hartley}
\label{subsec:hartley}

Another indication that someone was searching for the solution to the inverse problem before 1763 is a paragraph from David Hartley's book, {\it Observations on Man, His Frame, His Duty, and His Expectations}, published in 1749.
In a section where he discusses the refinements to the law of large numbers made by De Moivre, Hartley added a now famous paragraph.

\

\begin{center}
\begin{minipage}[c]{0.8\textwidth}
\begin{small}
An ingenious Friend has communicated to me a Solution of the inverse Problem, in which he has shewn what the Expectation is, when an Event has happened $p$ times, and failed $q$ times, that the original Ratio of the Causes for the Happening or Failing of an Event should deviate in any given Degree from that of $p$ to $q$. And it appears from this Solution, that where the Number of Trials is very great, the Deviation must be inconsiderable: Which shews that we may hope to determine the Proportions, and, by degrees, the whole Nature, of unknown Causes, by a sufficient Observation of their Effects.
\end{small}
\end{minipage}
\end{center}

\

The resemblance between the mentioned paragraph and the opening lines on Bayes's {\it Essay} is evident:

\

\begin{center}
\begin{minipage}[c]{0.85\textwidth}\small{
{\it Given} the number of times in which an unknown event has happened and failed: {\it Required} the chance that the probability of its happening in a single trial lies somewhere between any two degrees of probability that can be named.}
\end{minipage}
\end{center}

\

\citealp[p.~256]{daston1988} and \citealp[p.~331]{dale2003} comment on the resemblance, the later mentioning that Hartley used the term {\it expectation}, on which Bayes's definition of probability was based on, following the classic approach tradition of the early probabilists Pascal and Huygens.

The paragraph from \cite{hartley1749} led historians of probability and statistics to search for a personal connection between David Hartley and Thomas Bayes.\footnote{The first to conjecture about the connection was probably \cite{singer1979}. See also \cite{stigler1983, stigler2013, dale1986,dale2003}.} After the discovery of some of Bayes's unpublished manuscripts made by \cite{dale1986}, the connection was viewed as plausible, but no conclusive evidence was presented. See also \cite{bell2002, bell2007} and \cite{stigler2013} with respect to this possible connection. 
In \cite{allen1999}, Richard Allen, Hartley's biographer, also conjectures about a possible relationship, but did not present a conclusive answer.

Bayes and Hartley were both fellows of the Royal Society, Hartley elected in 1736 and Bayes in 1742, but apparently could not be considered individuals with mutual interests.
Hartley was a physician, son of an Anglican clergyman, while Bayes was a Nonconformist, a dissenting minister with an interest in mathematics.
However, looking closely, it seems quite plausible that they developed some sort of connection.
Their biographies \citep{allen1999, bell2004, dale2003} show their shared common interests, such as mathematical and theological subjects, and acquaintances.\footnote{William Whiston ($1667-1752$) and Isaac Maddox ($1697-1759$) to mention only two. Whiston, former Lucasian professor of mathematics was expelled from Cambridge in 1710 due to his religious views. See \citealp[p.~35]{allen1999} for the relationship of Whiston and Hartley and \citealp[p.~83]{dale2003} for his meeting with Bayes. Originally a dissenter, Maddox studied theology at the University of Edinburgh \citep{haydon2009}, where he and Bayes attended the logic and metaphysics course taught by Colin Drummond (c. $1685-1753$) in 1719. See the Matriculation Album of the University of Edinburgh. Names under the inscription {\it Discipuli Domini Colini Drummond qui vigesimo-septimo die Februarij, MDCCXIX Subscripserunt} and \cite{maddox1750} for his friendship with Hartley.}
Therefore, for the purposes of this work, we will assume that Bayes was indeed Hartley's ``ingenious Friend''.

We have been able to establish an early connection between Hartley and Bayes, albeit a brief one. The Royal Society of London, of which both Bayes and Hartley were fellows, had weekly meetings to discuss scientific reports from its members and others that were considered interesting.
The minutes of these meetings do not have a list with all those present, only of those who were not members and were invited by other fellows.
Those who brought some subject or text to be discussed were also mentioned.

It was usual for men interested in being elected fellows to attend at least one meeting under the sponsorship of another fellow, so that the candidate could be introduced to the other members of the Society.
On folio 380 of volume 17 of the Journal Books we read that Bayes attended a meeting of the Society on March 25, 1742, brought by John Belchier, a surgeon at Guy's Hospital in London.
The minutes of the same meeting continue on folio 381, where we find the following entry.

\

\begin{center}
\begin{minipage}[c]{0.8\textwidth}
\begin{small}
Doctor Hartley presented a pamphlet intitled, An account of the success of Mrs Stephens's medicines for the stone in the case of James Kirkpatrick Doctor of Divinity \& M.D. printed at Belfast in Ireland, 1739.
For which he had thanks.
\end{small}
\end{minipage}
\end{center}

\

David Hartley was a faithful advocate of Mrs. Stephens's medicine to treat bladder stones, having published several reports of successful cases of its use, including the one mentioned above.\footnote{See \cite{allen1999} for more on Hartley and the medicine for bladder stones.}
These minutes give us, so far, the only known mention of an occasion when Thomas Bayes and David Hartley were both present.

But why Hartley did not mention Bayes explicitly? 
Judging from his notes, letters and published works, we may say that Bayes was a quiet man who wanted to avoid quarrels. 
In 1747 he sent a letter to Philip Stanhope ($1714-1786$), second Earl  Stanhope, who studied mathematics under De Moivre.\footnote{Stanhope was one of the fellows who proposed Bayes's  election to the Royal Society in 1742.} The subject was a series firstly derived by James Stirling ($1692-1770$) and used by De Moivre in the demonstration of the normal approximation to binomial probabilities. 
De Moivre believed the series converged, providing an approximation to $\sqrt{2\pi}$. 
In the letter to Stanhope, Bayes showed the series diverges. 
He started the communication with “It has been asserted by some eminent mathematicians”, avoiding the reference of De Moivre and Stirling since he knew Stanhope would certainly inform the Frenchman about the mistake. The letter was published in the {\it Philosophical Transactions}, after Bayes's death.\footnote{\cite{bayes1763b}.}

He published only two pamphlets in his lifetime, both anonymously, both on controversial subjects: \cite{bayes1731} on divine benevolence, replying the views of John Balguy ($1686-1748$), an Anglican theologian, and \cite{bayes1736} on differential calculus, defending Newton’s approach against the criticism made by George Berkley ($1685-1753$), the bishop of Cloyne, in {\it The Analyst} (1734). Although his contemporaries knew that he was the author of both works,\footnote{See \cite{doddridge1763} and \cite{demorgan1860}. One of Bayes's obituaries says that he was the author of a well received pamphlet on divine benevolence. See {\it The Library Or Moral and Critical Magazine}, volume 1, May 1761, p.110.} they would not mention his authorship. Even \cite{price1758}, the first published book of his friend Richard Price, mentions \cite{bayes1731} but not the name of the author. Only the third edition of Price’s book, published in 1787, mentions Bayes explicitly. Thus, a plausible hypothesis is that Bayes asked Hartley to not be mentioned.

\section{Bayes and Hume}

We think Bayes's work on probability was not motivated as a response to \cite{hume1748}.
Firstly because in his publications and notes Bayes never sought applications, practical or philosophical, for the mathematical results he derived.
Secondly because, if Bayes communicated his findings to Hartley as we believe, the chronologies of the publications of \cite{hume1748} and \cite{hartley1749} indicate that it is unlikely that Bayes wanted to refute Hume.

Hume's book with the essay on miracles was advertised in April of 1748.\footnote{See \citep[p.~xxxvi]{hume2000}, the {\it Gentleman's Magazine}, volume 18, page 192, April of 1748 and {\it The London Magazine}, volume 17, page 240, May of 1748.}
\cite{stigler1983} says that Hartley's book was substantially complete by 1739 and fully finished by 1745 but \cite{webb1988} reports that Hartley prepared a new draft of {\it Observations on Man} that was completed only in the winter of 1747.
This draft was sent to some of his friends for corrections and useful remarks, and perhaps that was the time when Hartley's ``ingenious Friend'' communicated his discovery.
Corroborating this hypothesis, notice that the dates found on Bayes's notebook\footnote{\cite{dale1986}.} show that he was working on the {\it Essay} circa $1747-1750$.

The final version of \cite{hartley1749} was sent to the press in the beginning of December of 1748, and a presentation copy was sent to the Royal Society in February of 1749.\footnote{Journal Book Meetings, volume 21, folio 65, The Royal Society of London. See also the {\it Gentleman's Magazine}, volume 19, page 96, February of 1749.}
It is impossible to know when Hartley wrote the paragraph about the inverse problem but, if Bayes was Hartley's  friend and wanted to refute Hume, he had no more than eight months to read \cite{hume1748}, think about a reply, solve the problem and communicate the solution to Hartley, which seem unlikely.\footnote{On the other hand, since Hume's main point on induction was already present in \cite{hume1739} and \cite{hume1740}, it is possible that Bayes was already thinking about a reply before 1748 and all the debate about miracles raised by the {\it Enquiry} just gave him the final incentive to work on a rebuttal. 
Favoring the hypothesis that Bayes wanted to refute Hume's argument, \cite{stigler2013} found offprints of Bayes's essay with a different title: ``A Method of Calculating the Exact Probability of All Conclusions founded on Induction''.
A follow up of \cite{bayes1763a}, \cite{price1764}, published with the title ``A Demonstration of the Second Rule in the Essay toward the Solution of a Problem in the Doctrine of Chances'', also has an offprint with a different title, ``A  Supplement  to  the  Essay  on  a  Method of Calculating the Exact Probability of All Conclusions founded on Induction''.
Since the offprints were printed at the author's expense, \cite{stigler2013} says that Price almost certainly chose these titles, which are more informative and support better the case against Hume. This led Stigler to claim that it is not improbable that Bayes saw his work as a defensive tool to combat Hume, like Price did.}

\section{Bayes and De Moivre}

\cite{bellhouse2011} shows that Bayes did not study probability under De Moivre.
However, we can affirm that Bayes knew {\it The Doctrine of Chances}: letters exchanged between Bayes and Stanhope in the late 1740's\footnote{See \cite{bell2007} and \citealp[ch.~13]{bellhouse2011}.} show them discussing, among other mathematical subjects, probability problems found in the \cite{moivre1738}.

One topic of discussion in their correspondence was  the problem of runs: given a sequence of $n\in \mathbb{N}$ independent Bernoulli trials all with the same probability of success, what is the probability of a run or streak of $r\leq n$, $r\in\mathbb{N}$, successes in a row?\footnote{Problem LXXXVIII of \cite{moivre1738}.}
%De Moivre solved the problem using a generating function, without explaining how he derived it.
%Stanhope found an alternative but correct solution.
%Bayes's solution, sent to Stanhope, is wrong.
It is worth noticing that this is the first problem discussed after the translation of the {\it Approximatio} given in \cite{moivre1738}.\footnote{See \citealp[p.~243]{moivre1738}.}

Another topic discussed by Bayes and Stanhope was mentioned above: the series used to derive the normal approximation to binomial probabilities.
The result was published in the {\it Approximatio} and in \cite{moivre1738}.
In a manuscript sent to Stanhope, Bayes used a different argument to derive his own approximation to the factorial of a large positive integer, and then applied it to approximate the ratio of the middle term in the expansion of $(1+1)^n$ to $2^n$, ultimately obtaining the same approximation found by De Moivre.
If Stanhope informed De Moivre about Bayes's work, it would probably have annoyed the Frenchman.
Bayes's demonstration was an indication of an error, even though minor, but in a major discovery.

Hence, the Bayes$-$Stanhope correspondence shows that they were studying the final part of \cite{moivre1738} circa $1745-47$.
Bayes's notebook corroborates this hypothesis since it has notes on probability made between $1747$ and $1750$ with results published in his {\it Essay}.\footnote{See \cite{dale1986, dale2003}}
Our conjecture is that Bayes, while studying the {\it Approximatio} thought about a different approach to find inverse probabilities.
He eventually found an innovative way sometime before the end of 1748, being able to communicate the main point of his finding to Hartley.

%Studying De Moivre (1738) with Stanhope, Bayes found the mistake in De Moivre's series and solved the inverse problem c. 1747-1750 because he believed De Moivre's approach was not satisfactory. That was the "problem in the Doctrine of Chances".

%- previous quarrels: Bayes vs De Moivre (Bellhouse (2011)).

\section{Bayes and Murdoch}

%De Moivre and Murdoch became aware (via Stanhope) of Bayes's results sometime between 1747 and 1755. De Moivre didn't like it. Murdoch wrote Remark II to claim De Moivre's priority on the solution, reply Hume and tease Bayes.

%Thinking about Remark II, there are no other reasons for the text to be there. It doesn't add relevant theory, it doesn't add philosophical points (Remark I already made them) and it's unrelated to the problems following the text.

The third edition of {\it The Doctrine of Chances} was published in 1756, two years after De Moivre's death.
In the advertisement for the 1756 edition we read that De Moivre, due to his feeble health, trusted the new edition to ``one of his Friends; to whom he gave a Copy of the former [1738 edition], with some marginal Corrections and Additions, in his own hand writing. To these the Editor has added a few more, where they were thought necessary.''

The mentioned friend was Patrick Murdoch, a Scottish Anglican clergyman who was also recognized as a good mathematician.
In 1723 he graduated from the University of Edinburgh, where he was contemporary of Thomas Bayes, Nathaniel Carpenter (Bayes's cousin) and Isaac Maddox, who became bishop of Worcester and friend of David Hartley.
Murdoch, Maddox and Bayes attended the same logic course taught by Colin Drummond in 1719.\footnote{Matriculation Album of the University of Edinburgh.}
%Murdoch's biography mentions that he was ``pupil and friend''\footnote{See \cite{googwin2016}.} of Colin Maclaurin ($1698-1746$), but he did not study mathematics at Edinburgh under Maclaurin, who started to teach there only in 1725.
%Murdoch was elected fellow of the Royal Society in 1745.

De Moivre probably chose Murdoch to edit his book because the clergyman had a successful experience editing Maclaurin's papers after his death.
The editorial work resulted in {\it An Account of Sir Isaac Newton's Philosophical Discoveries, In Four Books} published in $1748$ with a prefatory biography of Maclaurin written by Murdoch.
%The book is considered ``one of the most adept popular expositions of Newtonian natural philosophy published in the Enlightenment.''\footnote{See \citealp[p.~102]{wood2003}.}
\cite{bellhouse2011} shows that Murdoch asked De Moivre's friends and former pupils, like Stanhope, to send unpublished material to be added in an appendix to the third edition, without naming the authors.
An analysis of Stanhope's unpublished works on probability shows that it is unlikely that he submitted anything to Murdoch.
Also according to \cite{bellhouse2011}, it is impossible to know which parts of the 1756 edition were added by Murdoch, Henry Stewart Stevens (d. 1760) and George Lewis Scott ($1708-1780$), who all helped to edit the book.
%Scott and Stevens studied mathematics under De Moivre. They both became lawyers and were elected fellows of the Royal Society, in 1737 and 1740 respectively, sponsored by their master. 

However, some parts of the added text can be plausibly attributed to hands that were not De Moivre's.
One of them is a remark added after the translation of the {\it Approximatio}.
The text after Corollary 10, already present in the 1738 edition, was named {\sc Remark I} and ended defending the design argument.
The text that follows is a more extensive defense of the same argument.
Named as {\sc Remark II}, it is almost three pages long, and starts with:\footnote{\citealp[p.~251]{moivre1756}.}

\

\begin{center}
\begin{minipage}[c]{0.8\textwidth}\small{
As, upon the Supposition of a certain determinate Law according to which any Event is to happen, we demonstrate that the Ratio of Happenings will continually approach to that Law, as the Experiments or Observations are multiplied: so, {\it conversely}, if from numberless Observations we find the Ratio of the Events to converge to a determinate quantity, as to the Ratio of $P$ to $Q$; then we conclude that this Ratio expresses the determinate Law according to which the Event is to happen.}
\end{minipage}
\end{center}

\

The text basically states that De Moivre's result given in the {\it Approximatio} would also provide a proof for the inverse problem, if we could perform ``numberless'' or an infinite number of observations. 
The following paragraph argues to prove the statement by contradiction.

The similarity between the opening paragraph of {\sc Remark II} and Hartley's comment on his friend's discovery is evident\footnote{See the paragraph from \cite{hartley1749} quoted on section \ref{subsec:hartley}.} but, so far, the chronology of the texts remained unnoticed or was misinterpreted.\footnote{\cite{daston1988} says that Hartley ``did not consider De Moivre's `remark' a sufficient guarantee of the inverse theorem, for he went on to mention an unnamed `ingenious Friend' who had shown him the solution to the `inverse Problem' [...]''}
Since {\sc Remark II} was not in the 1738 edition and Hartley's book was published in 1749, it is certain that Hartley did not see De Moivre's (or his editors') argument about the inverse problem.
The influence was, most probably, the opposite: Murdoch read Hartley's book and basically reproduced the point made by his ``ingenious Friend'', crediting De Moivre.\footnote{Hartley was no mathematician but had a solid background on the subject, being able to understand and apply mathematical tools in his works, like \cite{hartley1733}, a pamphlet on inoculation.
However, it is unlikely that he collaborated with De Moivre on any mathematical work, since there is no surviving evidence of such a partnership.}
Murdoch's library was sold in 1777 with the libraries of four other people.\footnote{Zachary Pearce ($1690-1774$), bishop of Bangor and F.R.S., William Lowndes ($1652-1724$), English politician, and two other reverends, Pawlet St. John and Barshaw.}
The catalogue of the libraries lists one copy of \cite{hartley1749}, although it does not mention whether or not it belonged to Murdoch. 

There are other elements allowing us to credit Murdoch for the authorship of {\sc Remark II}.
Firstly, to reinforce his point on design, the author mentions a probabilistic argument for divine providence from the ratio of male to female births, made by the Scottish physician John Arbuthnot ($1667-1735$) in 1710.\footnote{John Arbuthnot (1710). ``An argument for Divine Providence, taken from the constant regularity observed in the births of both sexes''. {\it Philosophical Transactions of the Royal Society of London}, 27, p. 186–190.}
Arbuthnot is described in \cite{moivre1756} as an ``excellent Person,'' which, based on an earlier comment by De Moivre, may seem strange.

In the preface of the {\it The Doctrine of Chances} written in 1717, De Moivre remembers the time he read Huygens's {\it De Ratiocinnis in Ludo Aleae} (1657) and its English translation ``done by a very ingenious Gentleman, who, tho' capable of carrying the matter a great deal farther, was contented to follow his Original; adding [...] some few things more.'' 
The mentioned translation of Huygens, anonymously published in 1692, was assigned to Arbuthnot.
De Moivre's reference is simultaneously polite and ironic, perhaps showing his resentment of the Scottish group which Arbuthnot was part of.\footnote{De Moivre was involved in a rough dispute with another Scottish physician, George Cheyne ($1672-1743$), friend of Arbuthnot and David Gregory ($1666-1708$). In a letter to Johann Bernoulli, De Moivre says of Gregory that ``It seems to me that he is suffering from a malady common to the Scotch.'' See \cite{walker1934}.}

A second piece of evidence pointing to Murdoch's authorship is a response to David Hume ($1711-1776$) present in the text of {\sc Remark II}.
Murdoch and Hume were close friends of Andrew Millar ($1705-1768)$, a Scottish publisher of several important British authors of the eighteenth century, including De Moivre and Hume.
%Hume's views on miracles, published in \cite{hume1748}, led to strong reactions from theologians and clergymen, especially in Britain.
Section VI of \cite{hume1748}, the essay {\it Of Probability}, starts with the following sentence.\footnote{See \citealp[p.~93]{hume1748}.}

\

\begin{center}
\begin{minipage}[c]{0.8\textwidth}\small{
Tho' there be no such Thing as {\it Chance} in the World; our Ignorance of the real Cause of any Event has the same Influence on the Understanding, and begets a like Species of Belief or Opinion.}
\end{minipage}
\end{center}

\

Moreover, the essay {\it Of Liberty and Necessity} has the sentence below.\footnote{\citealp[Sec.~VIII,~p~.151]{hume1748}.}

\

\begin{center}
\begin{minipage}[c]{0.8\textwidth}\small{
'Tis universally allow'd, that nothing exists without a Cause of its Existence, and that Chance, when strictly examined, is a mere negative Word, and means not any real Power, which has any where, a Being in Nature.}
\end{minipage}
\end{center}

\

To those statements, the author of {\sc Remark II} replied with the following.\footnote{\citealp[p.~13]{hacking1990} points out these references but incorrectly identifies the text as from \cite{moivre1738}, when in fact it was  available only in the 1756 edition. 
This mistake led Hacking to assume that \cite{hume1748} was referring to De Moivre's epithet, that chance is a mere word, when the actual timing, and possible influence or reference, was the opposite.}

\

\begin{center}
\begin{minipage}[c]{0.8\textwidth}\small{
{\it Chance}, as we understand it, supposes the {\it Existence} of things, and their general known {\it Properties}: that a number of Dice, for instance, being thrown, each of them shall settle upon one or other of its Bases. After which, the {\it Probability} of an assigned Chance, that is of some particular disposition of the Dice, becomes as proper a subject of Investigation as any other quantity or Ratio can be.

But {\it Chance}, in atheistical writings or discourse, is a sound utterly insignificant: It imports no determination to any {\it mode of Existence}; nor indeed to {\it Existence} itself, more than to {\it non-existence}; it can neither be defined nor understood: nor can any Proposition concerning it be either affirmed or denied, excepting this one, `That it is a mere word.'}
\end{minipage}
\end{center}

\

Therefore, considering the length of {\sc Remark II}, its arduous defense of the design argument with complimentary references to Arbuthnot and Nicholas Bernoulli and a rebuttal of a sceptical philosopher,\footnote{Interestingly, Hume used at least three passages from \cite{mclaurin1748}, edited by Murdoch, in his works. Compare \citealp[p.~56]{hume1748} with \citealp[p.~105]{mclaurin1748} and \citealp[p.~42-44,45-46]{hume1779} with \citealp[p.381,378-9]{mclaurin1748}. Hume's library had a copy of \cite{mclaurin1748}, as showed by \cite{norton1996}.} we can safely conclude that the text was not penned by De Moivre.
On the other hand, it is plausible to assume it coming from a clergyman like Murdoch.

%Did Murdoch know about the discovery made by Hartley's friend?
%Once more, the text of {\sc Remark II} may help us to answer this question.
Finally, the last paragraph of the remark says that Bernoulli's result can be applied to several fields, as shown in Part IV of \cite{bernoulli1713} and the last sentence of the section is the following.

\

\begin{center}
\begin{minipage}[c]{0.8\textwidth}\small{
Yet there are Writers, of a Class indeed very different from that of {\it James Bernoulli}, who insinuate as if the {\it Doctrine of Probabilities} could have no place in any serious Enquiry; and that Studies of this kind, trivial and easy as they be, rather disqualify a man for reasoning on every other subject. Let the Reader chuse.}
\end{minipage}
\end{center}

\

This may be a reference to the closing paragraph of \cite{bayes1736}, the pamphlet on differential calculus.

\

\begin{center}
\begin{minipage}[c]{0.8\textwidth}\small{
To conclude; as I would not be thought, by any thing I have said, to be an enemy to true Logic and sound Metaphysics; and on the contrary think the most general use of the Mathematics is to inure us to a just way of thinking and arguing; it is a proper inquiry, I imagine, for those who have the direction of the education of youth, in what manner mathematical studies may be so pursued as most surely to answer this end [...] for so far as Mathematics do not tend to make men more sober and rational thinkers, wiser and better men, they are only to be considered as an amusement, which ought not to take us off from serious business.}
\end{minipage}
\end{center}

\

%\cite{bayes1736} was an anonymous pamphlet published to defend Newton's approach to calculus against the criticism made by George Berkeley ($1685-1753$), the bishop of Cloyne, in {\it The Analyst} (1734).
The text is probably informing Bayes's opinion on the purpose of teaching pure mathematics in dissenting academies.
\cite{bell2010} describes the pedagogical reasons that justified the teaching of mathematics in those academies.\footnote{Consider for instance the view of Philip Doddridge ($1702-1751$), a dissenting educator: he saw mathematics not only as a precursor to the study of the new Newtonian science, but also as a tool to help mankind to better understand how God works in creation.}

We do not have much evidence to judge Murdoch's relationship with Bayes, but a letter sent by Bayes to Stanhope in 1755 reveals something.\footnote{See \cite{bell2002}.} 
The letter shows Bayes commenting on an article written by Murdoch, probably at Stanhope's request, and the comments were not favourable.
If Bayes and Murdoch were close friends, the latter would ask directly for comments on his work, not waiting for someone else to be an intermediary in the request.

Therefore, we think it is plausible to assume that sometime between 1747 and 1755, Stanhope informed De Moivre and/or Murdoch about Bayes's results on inverse probability, without providing much detail.
De Moivre did not like it and Murdoch, asked by De Moivre or by his own initiative, wrote {\sc Remark II} to claim De Moivre's priority on the solution and to reply to both Hume and Bayes.
Considering {\sc Remark II} on its context, we realized there are no significant reasons for the text to be there: it does not add relevant theory or philosophical points ({\sc Remark I} already made them) and it is unrelated to the problems following the text.
Given this, we may reasonably conjecture that, in fact, Murdoch knew that Bayes was Hartley's ingenious friend, but considered De Moivre to be the first to have solved the inverse problem.
Richard Price, Bayes's friend and intellectual executor disagreed.

\section{Concluding remarks}

%Price replied in 1763: Bayes, not De Moivre, was the first to solve the problem (satisfactorily).

After Thomas Bayes's death his family asked Price to look into his papers, among which Price found notes which he edited and expanded in an appendix with numerical illustrations. 
The final result, named {\it An Essay Towards Solving a Problem in the Doctrine of Chances}, was sent with a cover letter to John Canton ($1718-1772$), a fellow of the Royal Society and common friend.
%The letter not only describes what Bayes have done %but also stands up for his efforts.

The first remark made by Price is that Bayes realized he needed what we now call a prior distribution in order to solve the inverse problem, something De Moivre and Jakob Bernoulli did not think of.
It is also clear from the text that Price rejected the alleged proof of the inverse problem presented in {\sc Remark II} of \cite{moivre1756},\footnote{Price used the 1756 edition as reference, so he knew {\sc Remark II}. Price mentioned that Bayes deduced the solution of the ``converse problem'', echoing \citealp[p.~251]{moivre1756} where we read, ``so, {\it conversely}''.} recognizing Bayes as the first to solve it.

\

\begin{center}
\begin{minipage}[c]{0.85\textwidth}\small{
It may be safely added, I fancy, that it is also a problem that has never before been solved. Mr. De Moivre, [...] has [...], after Bernoulli,[...], given rules to find the probability there is, that if a very great number of trials be made concerning any event, the proportion of the number of times it will happen, to the number of times it will fail in those trials, should differ less than by small assigned limits from the proportion of the probability of its happening to the probability of its failing in one single trial. But I know of no person who has shewn how to deduce the solution of the converse problem to this; namely, ``the number of times an unknown event has happened and failed being given, to find the chance that the probability of its happening should lie somewhere between any two named degrees of probability.'' What Mr. De Moivre has done therefore cannot be thought sufficient to make the consideration of this point unnecessary: especially, as the rules he has given are not pretended to be rigorously exact, except on supposition that the number of trials made are infinite; from whence it is not obvious how large the number of trials must be in order to make them exact enough to be depended on in practice.}
\end{minipage}
\end{center}

\

Price recognized De Moivre's work refining Bernoulli's result, but saw {\sc Remark II} as useless in practice, since it relies on the ``supposition that the number of trials made are infinite.''
Possibly defending his friend against the criticism of Murdoch, Price underlined the importance of the problem solved by Bayes.

\

\begin{center}
\begin{minipage}[c]{0.85\textwidth}\small{
Every judicious person will be sensible that the problem now mentioned is by no means merely a curious speculation in the doctrine of chances, but necessary to be solved in order to a sure foundation for all our reasonings concerning past facts, and what is likely to be hereafter.[...]

These observations prove that the problem enquired after in this essay is no less important than it is curious. 
}
\end{minipage}
\end{center}

\

The possibility of a defense of \cite{bayes1736} gains weight when Price explicitly mentions Murdoch's criticism against it.

\ 

\begin{center}
\begin{minipage}[c]{0.85\textwidth}\small{
Mr. De Moivre calls the problem he has thus solved, the hardest that can be proposed on the subject of chance. His solution he has applied to a very important purpose, and thereby shewn that those a remuch mistaken who have insinuated that the Doctrine of Chances in mathematics is of trivial consequence, and cannot have a place in any serious enquiry. The purpose I mean is [...] to confirm the argument taken from final causes for the existence of the Deity. It will be easy to see that the converse problem solved in this essay is more directly applicable to this purpose [...]
}
\end{minipage}
\end{center}

\

In a very polite way Price acknowledged De Moivre's (or Murdoch's?) efforts in trying to prove the design argument from probabilistic arguments but points out that Bayes's solution is more appropriate to this end.

\bibliography{mybiblio}
\bibliographystyle{apalike}

\section*{Appendix: Bernoulli's theorem and its inverse}

The objective of this appendix is to formally differentiate Bayes's theorem from the inverse use of Bernoulli's law of large numbers.
In order to do this we will use modern notation.
%According to \cite{matmuller2014}, Jakob Bernoulli stated and proved the first form of the weak Law of Large Numbers sometime between 1685 and 1689.\footnote{Apparently Bernoulli never called this result the ``golden theorem'', as some authors think. His derivation of the equation for a curve known as {\it elastica} was named by him as ``golden theorem''. See \cite{bernoulli1694} for the true ``aureum theorema''.}
If $\{X_i\}_{i=1}^{n}$, $n\in\mathbb{N}$, 
are independent Bernoulli random variables such that $E(X_i) = P(X_i=1)=\theta\in[0,1]$, then for any given real number $\varepsilon>0$, Jakob Bernoulli proved that:

\[
\lim_{n\to\infty}P(\theta-\varepsilon<\overline{X}_n<\theta+ \varepsilon\mid \theta)=1,
\]

\noindent
where $\overline{X}_n=\sum_{i=1}^nX_i/n$.
Notice that Bernoulli assumed that $\theta$ and $\varepsilon$ were known to compute $n$ such that $P(\theta-\varepsilon<\overline{X}_n<\theta+ \varepsilon\mid \theta)$ would be as close to 1 as desired.\footnote{
As a numerical example, \citealp[p.~238]{bernoulli1713} found $n=25,550$ for $\theta=3/5$ and $\varepsilon=1/50$ in order to make $P(|\theta-\overline{X}_n|<\varepsilon)=0.999$.}
Thus, the question Bernoulli answered was: assuming that the probability of an event is known, how many times an experiment ought to be repeated before one can be ``morally'' certain that the relative frequency of the event is close enough to the true probability?

If the event probability ($\theta$) is unknown and a given number of successes and failures is obtained after $n$ trials, one can use Bernoulli's law inversely by taking the relative frequency of successes as an approximation to the true but unknown probability. 
Although \cite{dale1988}, \cite{hacking1975} and \cite{todhunter1865} argue that Bernoulli had this use in mind, this view cannot be proven. 

On his side, Bayes wanted to solve the inverse problem as stated in Section 2: his goal was to compute for any given limits $\ell_1$ and $\ell_2$, $
P(\ell_1\leq\theta\leq\ell_2 ~ | ~ \overline{x}_n)$,
where $\overline{x}_n$ is the observed value of $\overline{X}_n$.
In fact, Bayes's result is not a limit theorem and, as such, cannot be formally seen as the inverse of Bernoulli's law.
\cite{laplace1774} eventually proved the inverse of Bernoulli's law based on inverse probability and his method of asymptotic approximation for integrals.

As put by \citealp[p.~140]{gorroochurn2012}, the difference between Bernoulli's law, the inverse use of Bernoulli's law and Bayes's theorem may be summarized as: 

\begin{itemize}
\item Bernoulli's law: for given $\theta$ and $\varepsilon>0,$ we can find $n\in\mathbb{N}$ such that
$P(\theta-\varepsilon<\overline{X}_{n}<\theta+\varepsilon \mid \theta)$ is arbitrarily close to 1;

\item inverse use of Bernoulli's law: when $\theta$ is unknown, given $\overline{x}_{n}$, $P\left(\overline{x}_n-\varepsilon<\theta<\overline{x}_n+\varepsilon\right)$ is close to $1$, when $n$ is large, so that $\overline{x}_{n}\approx \theta$.

\item Bayes's theorem: when $\theta$ is unknown, for given $\ell_{1}, \ell_{2}, \overline{x}_{n}$ and assuming that $\theta$ has some prior distribution, then we can calculate $P\left(\ell_{1}<\theta<\ell_{2} \mid \overline{x}_{n}\right)$.
\end{itemize}

Under this perspective, one may say that Bayes's approach was the ``correct'' one to solve the inverse problem as considered in this text.
The inverse use of Bernoulli's law by other authors, such as \cite{halley1693}, did not foreshadow by any way the method applied by Bayes to compute the required conditional probability.

\end{document}